\documentclass{svproc}
\pdfoutput=1
\usepackage{graphicx}
\usepackage{multicol}
\usepackage{footmisc}
\usepackage{amsmath}
\usepackage{subfigure}
	
\usepackage{soul}
\begin{document}
\mainmatter              
\title{Tuning Path Tracking Controllers for Autonomous Cars Using Reinforcement Learning}
\titlerunning{Path Tracking Controller Tuning Using Reinforcement Learning}  
%
\author{Ana Vilaça Carrasco\inst{1} \and João Silva Sequeira\inst{1,2} }
\authorrunning{Ana Carrasco et al.} 
%
\tocauthor{Ana Carrasco, João Silva Sequeira}
\institute{Instituto Superior Tecnico, University of Lisbon, Portugal
\and
Institute for Systems and Robotics, Portugal
\\
\{ana.vilaca.c, joao.silva.sequeira\}@tecnico.ulisboa.pt}

\maketitle              

\begin{abstract}
This paper proposes an adaptable path tracking control system based on Reinforcement Learning (RL) for autonomous cars.
A four-parameter controller shapes the behavior of the vehicle to navigate on lane changes and roundabouts.
The tuning of the tracker uses an \textit{educated} Q-Learning algorithm to minimize the lateral and steering trajectory errors.

The CARLA simulation environment was used both for training and testing.
The results show the vehicle is able to adapt its behavior to the different types of reference
trajectories, navigating safely with low tracking errors. The use of a ROS bridge between the CARLA and
the tracker results (i) in a realistic system, and (ii) simplifies the replacement of the CARLA
by a real vehicle.

An argument on the dependability of the overall architecture based on stability results
of non-smooth systems is presented at the end of the paper.
\end{abstract}

\section{Introduction}
\label{sec:intro}
Over the last decades, autonomous vehicles have quickly become a popular research subject, and are likely to have a significant societal impact, e.g., reducing accidents and traffic congestion, optimizing energy use, and have a more Eco-friendly impact in the world.

The architecture of an autonomous vehicle is built around the standard Guidance-Navigation and Control (GNC) structure.
The control of the vehicle (sometimes named path tracking) is responsible to follow a reference path accurately, ensuring the system has a stable and safe behavior, is robust to disturbances and can adapt to different environments.

The literature describes multiple path tracking control methods~\cite{YaoQ,PendletonS,SorniottiA}: Pure Pursuit and Stanley methods, conventional feedback controllers (ex. Linear Quadratic Regulators (LQR), Proportional Integral Derivative (PID) controllers), Iterative Learning Control (ILC), and Model Predictive Control (MPC), the most popular being PID controllers and MPC based controllers. Less conventional control structures are also proposed to tackle problems such as non-linearity, parameter uncertainties and external disturbances, like $H_{\infty}$ controllers and Sliding Mode Controllers (SMC).

Machine Learning (ML) offers many benefits to the field of Autonomous Driving Systems (ADS): self-optimization based on collected data and new environments, the ability to specify the desired behavior of the system, and increased generalization capacity. When designing an intelligent autonomous vehicle, a popular approach is a Supervised Learning (SL) based end-to-end architecture with a complex Neural Network (NN)~\cite{Bojarski2016EndCars}.
However, these solutions generally have great computational complexity during training and end-to-end architectures are associated with the ``black-box'' problem~\cite{Kuutti}.
Alternatively, combining model-based controllers with learning algorithms can preserve the properties and methodologies from traditional controller design and analysis but also provides robustness and adaptability from the learning component.
Moreover, by carefully constraining the parameter space explored during the learning phase(s), one obtains dependable by construction architectures.

Learning techniques have been used in conjunction with MPC to improve path tracking~\cite{Brunner,OstafewMPC}. However, Deep Learning (DL) based control models have been shown to outperform these more popular methods~\cite{DongLi,YunxiaoShan}. Despite the popularity of advanced control techniques, simple and fine-tuned feedback path tracking controllers can provide good performance for a variety of conditions~\cite{SorniottiA}.

A control architecture combining traditional path tracking controllers (pure pursuit and PID) with Reinforcement Learning modules, was proposed in~\cite{ChenMing,YunxiaoShan,LongshengChen},
and reported  to effectively improve the performance of the traditional controllers.
Using RL algorithms to optimize the parameters of PID controllers has also been proposed~\cite{AhmedSA,Kofinas2018FuzzyControl,Shi2018AdaptiveAlgorithm}. 

The control modules of autonomous vehicles in the literature have been validated in a variety of maneuvers, including lane keeping, lane change, ramp merging and intersection navigation~\cite{Farazi}. 

The present work describes a path tracking controller using a Reinforcement Learning (RL) agent to perform offline parameter tuning. The RL agent, using a discretized tabular variation of the Q-Learning algorithm~\cite{BartoSutton}, is trained to fine-tune the controller's gains while performing the lane change and roundabout maneuvers.

Similar maneuvers (or path types) are used in \cite{KCKoh} to validate their path tracking method.
Their work also compares its proposed method to a ``conventional'' tracking method that is very
similar to the four-parameter controller used in this work (yet missing the learning component).
  
A Q-Learning based framework for longitudinal and lateral control was proposed in \cite{WangP} and experiments showing the system while performing a lane change maneuver are presented. 


The proposed architecture also includes a module that identifies when to perform each maneuver and changes the gains appropriately, and a safety watchdog module that controls the vehicle's velocity.

\section{Implementation}
\label{sec:implementation}

\begin{figure}
  \centering
  \includegraphics[width=0.70\textwidth]{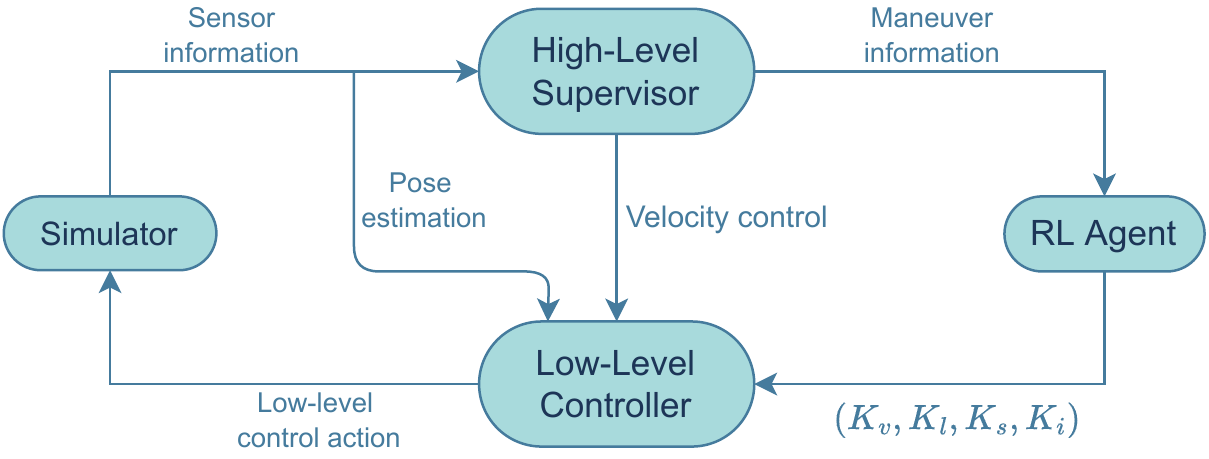}
  \caption{Full system architecture}
  \label{fig:full_arch}
\end{figure}

The proposed architecture is presented in Figure \ref{fig:full_arch}.
The simulator implements a regular vehicle with multiple sensors attached.
The use of the ROS framework to bridge vehicle and the overall control architecture allows a quick replacement
of the simulator by a real vehicle.
The low-level controller drives the vehicle through a predefined reference path by calculating and imposing values for velocities and steering angle. 
The RL agent is trained to find the best set of gains for each maneuver.
The maneuvers tested were lane changing (to the right) in a straight road and circulating in a roundabout. 

The high-level supervisor monitors if the linear velocity is within the imposed limits and if the vehicle is required to perform one of the maneuvers. If so, it sends that information to the RL agent, which will then set the gains to the appropriate fine-tuned values for that maneuver. Those fine-tuned gains, denoted $(K_v, K_l, K_s, K_i)$, are then sent to the low-level controller to calculate the steering angle, $\phi$, the linear and angular velocities, $v$ and $w_s$ \textendash~ these three values define a low-level control action. If necessary, the value for $v$ is overridden by the high-level supervisor, to stay within limits. The simulator also communicates directly with the low-level controller, sending an estimation of the vehicle's current pose, based solely on the vehicle's odometry.


\subsection{Low-level Controller}
\label{sec:LLcontroller}

The low-level control module controls the trajectory of the vehicle by adjusting the values of the steering angle $\phi$, linear velocity, $v$, and angular velocity, $\omega_s$,  in real-time, with the goal of minimizing the error between the reference and the actual pose of the vehicle. The control laws,  which are based in a nonholonomic vehicle model, are a function of the error between the reference and the actual pose, in the vehicle frame,
$^be$. The error in the world frame, $^we$, is

\begin{equation}
    ^we = [x_{ref} - x, y_{ref} - y, \theta_{ref}-\theta]\;, 
\label{eq:e_w}
\end{equation}

\noindent
where $(x_{ref},y_{ref},\theta_{ref})$ is the reference pose and $(x,y,\theta)$ is the vehicle's current pose. The control laws to yield $v$, $\omega_s$ and $\phi$, written at discrete time $k$, are given by,

\begin{align}
  v_k &= \pmb{K_v}\, ^be_{x_k},   &   \omega_{s_k} &= \pmb{K_s}\, ^be_{\theta_k} + \pmb{K_l}\, ^be_{y_k},    &   \phi_k &= \pmb{K_i}\, \phi_{k-1} + \pmb{K_i}\,h\, \omega_{s_k}, \label{eq_phi} 
\end{align}

\noindent
where $h$ is a time step and the $^be$ is obtained from $^we$ by means of a rotation matrix of a $\theta^{\circ}$ rotation around the Z axis.
The last equation from \eqref{eq_phi} is a low-pass filter that removes
unwanted fast changes in $\omega_s$.

\begin{figure}[h]
\centering
\includegraphics[width=0.9\textwidth]{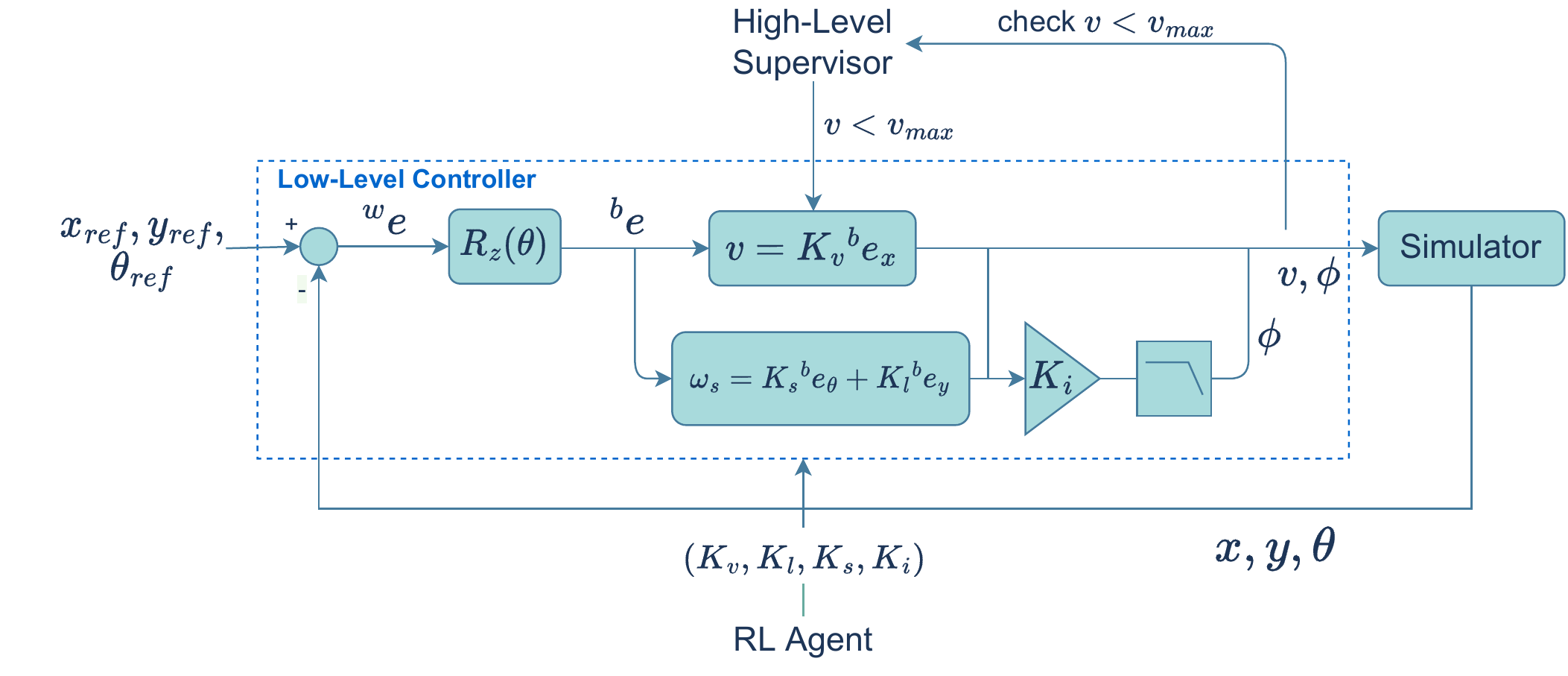}
\caption{Diagram of the low-level controller.}
\label{fig:controller_arch}    
\end{figure}

The $v$, $\omega_s$ and $\phi$ are then converted into control actions and sent to the vehicle. The trajectory controller gains are the linear velocity gain, $K_v$, the steering gain, $K_s$, the linear gain, $K_l$, and the lowpass filter gain, $K_i$ (see Figure \ref{fig:controller_arch}).

The loop iterates until the destination is reached (i.e., if the current position is close enough to the last position in the path), a collision is registered or the simulation time ends.

\subsection{Simulator}
\label{sec:sim}

CARLA~\cite{CARLA} is an open-source simulator designed for research on autonomous driving~\cite{LongshengChen,samak2021proximally,YunxiaoShan}. It simulates urban realistic environments (in terms of rendering and physics).
A ROS bridge allows direct communication with the simulated vehicle, through publishers and subscribers, and also provides a way to customize the vehicle setup. A ``Tesla Model 3'' vehicle was chosen, including  speedometer,  collision detector, and odometry sensors.

\begin{figure}[h]
    \centering
    \includegraphics[width=0.65\textwidth]{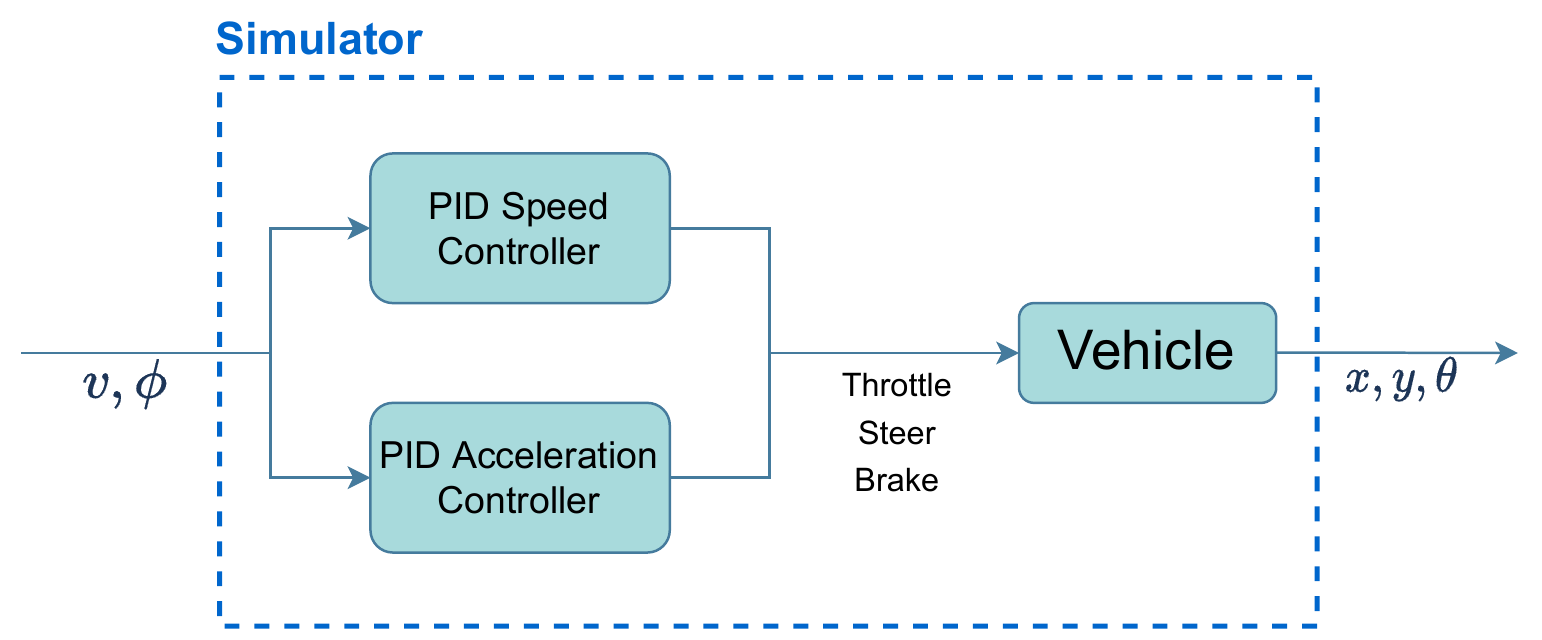}
    \caption{Diagram of the vehicle simulator.}
    \label{fig:sim}
\end{figure}

In this project, the simulator runs with a fixed time-step (the time span between two simulation frames) of 0.01 (simulation) seconds. Figure \ref{fig:sim} illustrates, in a simple way, how the vehicle simulator transforms the linear velocity ($v$) and steering angle ($\phi$) provided by the low-level controller into messages that control the throttle, steer and brake values of the vehicle. The current pose of the vehicle is updated by subscribing to an odometry publisher provided by the simulator.

\subsection{High-level Supervisor}
\label{sec:HLcontroller}

\begin{figure*}[h]
  \centering
  \includegraphics[width=0.8\textwidth]{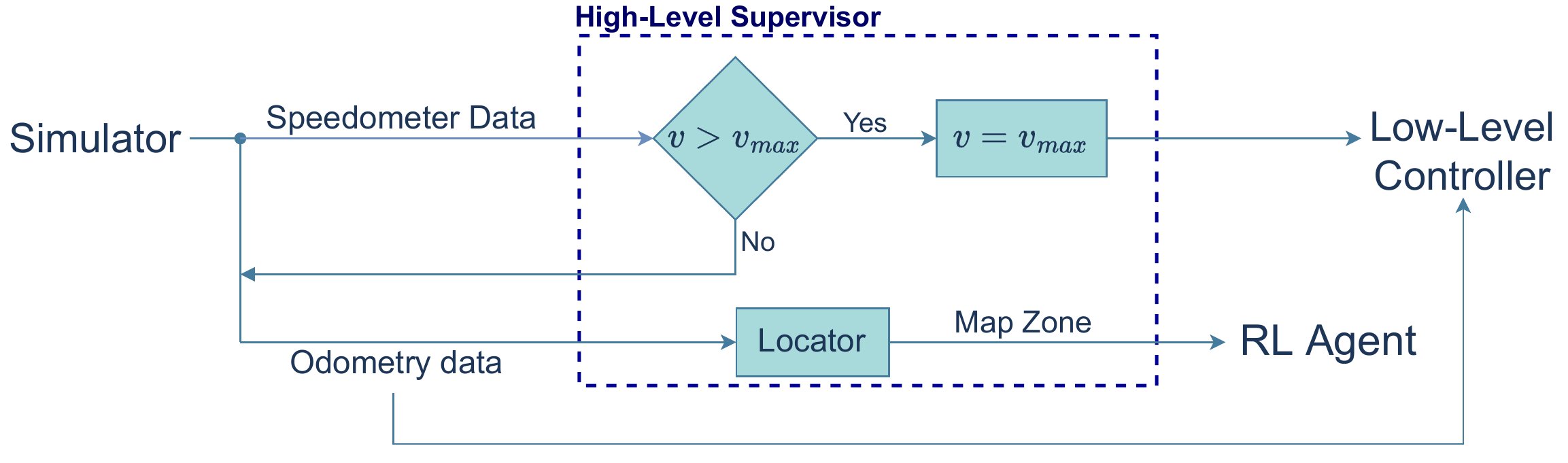}
  \caption{Diagram of the high-level supervisor's internal operations.}
  \label{fig:HL_controller}
\end{figure*}

The high-level Supervisor works as both an event manager and a safety module (see Figure \ref{fig:HL_controller}). It determines if the vehicle needs to perform one of the two maneuvers based on which zone of the map the vehicle is currently in. It also enforces a speed limit, overriding, if necessary, the linear velocity value calculated by the low-level controller.

In the experiments performed, the map was divided into zones, each of which associated with an event (Figure \ref{fig:town03_zones}). In the blue zone the vehicle performs a lane change and in the red zone, it navigates a roundabout. The reference path is shown as the black line.

\vspace{-7mm}
\subsection{Reinforcement Learning Agent}
\label{sec:RL}

The RL agent is responsible for tuning the trajectory controller gains using a variation of the Q-Learning algorithm. In the original Q-Learning, a discretized Q-Table takes an interval of gains and finds the values with which the vehicle presents the best performance. 
In the variation used in the paper, referred as \textit{educated Q-Learning}, this interval of gains is narrowed down to the most chosen values throughout the training. This facilitates the selection of the best gains by deliberately reducing the action space the algorithm has to explore. Performance evaluation is translated in the reward function of the algorithm.
The RL environment is defined as follows:

\textbf{States:} An array with the average of the absolute value of the lateral and orientation errors, $S = [E_y, E_\theta]$. Each of the error values have low and high limits, $E_{LOW}$ and $E_{HIGH}$, and are discretized into 40 units;

\textbf{Actions:} Each action, $A = [a_0, a_1, a_2, a_3]$, is represented by an array. 
There are 81 different actions. The gains are adjusted by the action array through the following expressions,
\begin{align}
K_v & = K_v + h_0a_0,  &   K_l = K_l + h_1a_1, \nonumber \\ 
K_s & = K_s + h_2a_2,  &   K_i = K_i + h_3a_3. \label{eq:gains}
\end{align}

The $a_0$, $a_1$, $a_2$ and $a_3$ can take the values 1, 0 or -1. The values $h_0$, $h_1$, $h_2$ and $h_3$ are positive constants that will either be ignored, subtracted or added to the previous value of the gain.\par

\textbf{Terminal condition: }Given the difficulty reaching state $S_0 = [0 , 0]$, the adopted approach was to consider any state that would come closer to $S_0$ to be the terminal state. As a result, if the current state is closer to $S_0$ than the closest state recorded so far, then a terminal state was reached. To determine the distance of a state to the state $S_0$, $d(S,S_0)$, the algorithm uses a weighted euclidean distance. Since the lateral error values, $E_y$, are generally 10 times greater than the orientation error values, $E_{\theta}$, the weight array used was $[1, 10]$:
\begin{equation}
     d(S,S_0) = \sqrt{|E_y|^{2}+10\times|E_{\theta}|^{2}}
\end{equation}

The sets of gains that produce the terminal states are referred to as the \textbf{terminal gains}. If, for the last 5 terminal sets of gains, a gain has a constant value, then that gain's range is locked into that value for the rest of the training \textendash~this defines the \textit{educated Q-Learning} variation presented in this paper.\par

\textbf{Reward function: }The reward function chosen for this work is defined by the equation,
\begin{equation}
    R = \frac{1}{1+d(S', S_0)} - \frac{1}{1+d(S, S_0)}\;,
\label{eq:reward_f}
\end{equation}
\noindent
where $d(S', S_0)$ is the distance between the new state $S'$ and $S_0$ and $d(S, S_0)$ is the distance between the current state $S$ and $S_0$. 
\begin{figure}
\scriptsize
\centering
\includegraphics[width=1\textwidth]{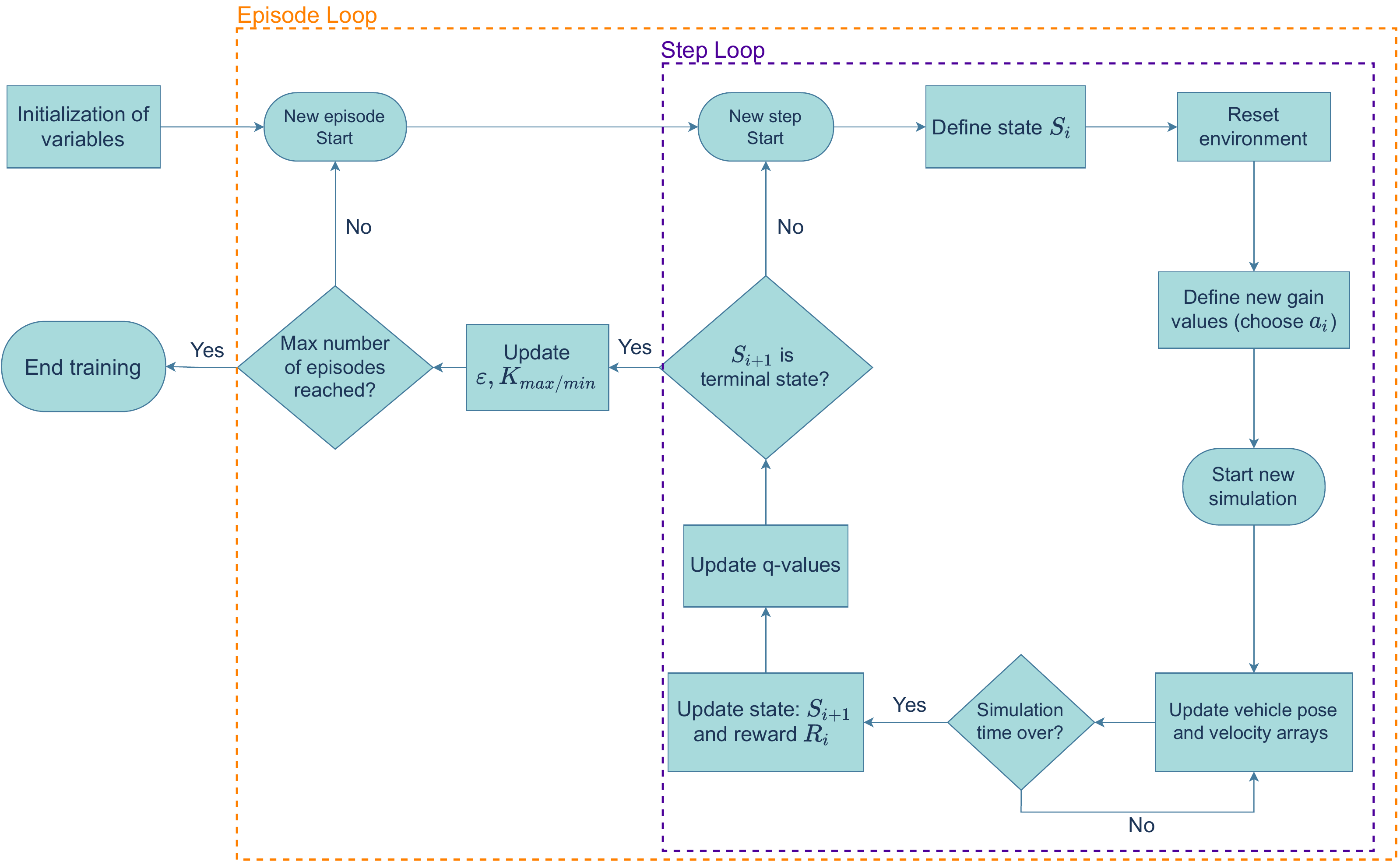}
\caption{Diagram of the training algorithm.}
\label{fig:QL_alg}
\end{figure}

\noindent
This function is based on the one used in \cite{Kofinas2018FuzzyControl}.
Also, if a collision is registered the reward is decreased by a defined value.

\textbf{Training Algorithm:} The RL agent was trained to perform two different maneuvers: a lane changing maneuver in a straight road and driving in a roundabout. The algorithm used to train the RL agent is shown by the diagram in Figure \ref{fig:QL_alg}. 
The agent was trained over a certain number of episodes, each of which is divided by steps. Each step, a current state, $S_i$, is defined based on the last error average registered. Then, an action is taken and the new gains are defined, after which a new simulation starts, with the system's controller guiding the vehicle through the reference path. After the simulation stops, the new state, $S_{i+1}$, and the reward, $R_i$, are updated. With these values, the Q-Table is updated based on equation (6.8) in \cite{BartoSutton}(p.131). If the new state, $S_{i+1}$ does not satisfy the terminal condition, this cycle repeats in a new step. Otherwise, the episode ends, \textbf{$\pmb{\varepsilon}$} and the new gain range is updated (based on the \textit{educated Q-Learning} variation) and a new episode starts.

\section{Simulation Results}
\label{sec:results}

The values chosen for the parameters, for each of the maneuvers, are shown in the table below.

\begin{table}[h!]
\caption{Parameter values for each of the training environments,where $n$ is the number of episodes.}
\begin{center}
\begin{tabular}{c@{\quad}cc}
\hline
\multicolumn{1}{c}{\rule{0pt}{12pt}\textbf{Variable} }&\multicolumn{1}{c}{\textbf{Lane Change}}&\multicolumn{1}{c}{\textbf{Roundabout}}\\[2pt]
\hline\rule{0pt}{12pt}
\textbf{Loop time} & 5 & 30\\
\textbf{$\pmb{\gamma}$} & 0.9 & 0.9\\
\textbf{$\pmb{E_{HIGH}}$ (m)} & {[}3 , 0.4] & {[}1 , 0.1]\\
\textbf{$\pmb{E_{LOW}}$ (m)} & {[}0 , 0] & {[}0 , 0]\\
\textbf{$\pmb{K_{min} (K_v, K_l, K_s, K_i)}$} & {[}0.1, 1, 1, 0.7] & {[}1, 1, 1, 0.7]\\
\textbf{$\pmb{K_{max}(K_v, K_l, K_s, K_i)}$} & {[}3, 21, 21, 0.98] & {[}5.8, 21, 21, 0.98]\\
\textbf{$\pmb{[h_0, h_1, h_2, h_3]}$} & {[}0.58, 5, 5, 0.07] & {[}1.2, 5, 5, 0.07]\\
\textbf{$\pmb{\phi}$ range ($^\circ$)} & $\pm 30$ & $\pm 30$\\
\textbf{$\pmb{\varepsilon}$} (start) & $1$ & $1$\\
\textbf{$\pmb{\varepsilon}$ decay} & $1/\text{(n/2)}$ & $1/\text{(n/2)}$\\
\textbf{Step Limit} & $130$ & $100$\\[2pt]
\hline
\end{tabular}
\end{center}
\label{tab:params}
\end{table}

The \textbf{Loop time} represents the time of each training step, in simulated seconds. $\pmb{\gamma}$ is the discount factor used for the Q-Learning equation used. $\pmb{E_{HIGH/LOW}}$ are the state limits (see Section \ref{sec:RL}). $\pmb{K_{max/min}}$ define the range of gains explored. The values $\pmb{[h_0, h_1, h_2, h_3]}$ are the positive constants that define the action values (eq. \eqref{eq:gains}). \textbf{$\pmb{\varepsilon}$ decay} and start value refer to the $\pmb{\varepsilon}$-greedy policy \cite{BartoSutton} used in the Q-Learning algorithm. $\pmb{\phi}$ \textbf{range} is the default steering angle range used during training. \textbf{Step limit} refers to the maximum number of steps an episode can have, a condition that prevents unfeasible training times. 

\begin{figure}[h]
\centering
       \subfigure{\includegraphics[width=0.7\textwidth]{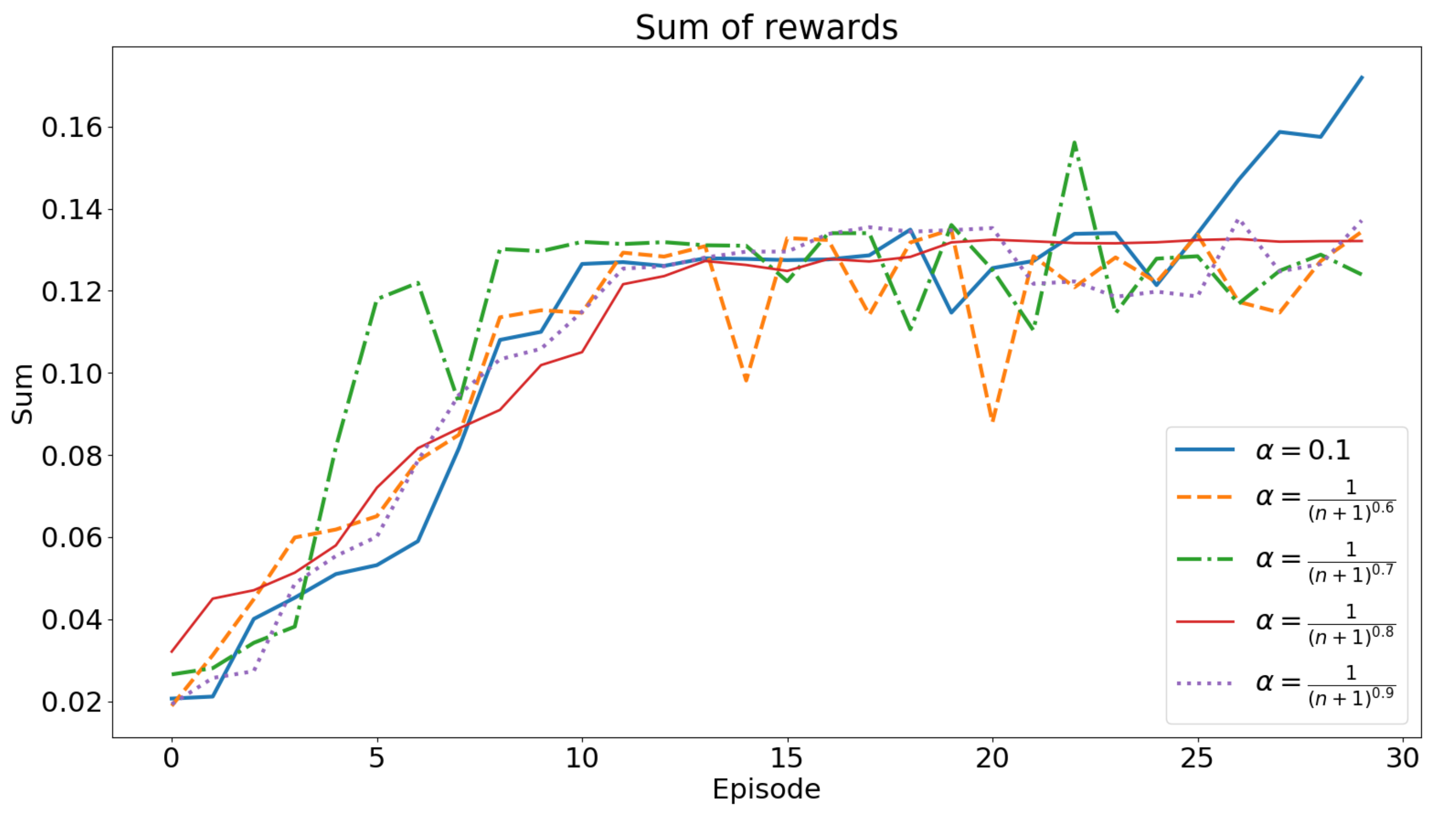}}
       \caption{Lane changing training: Sum of rewards per episode.}
       \label{fig:test_straight_sum}
       \centering
       \subfigure{\includegraphics[width=0.7\textwidth]{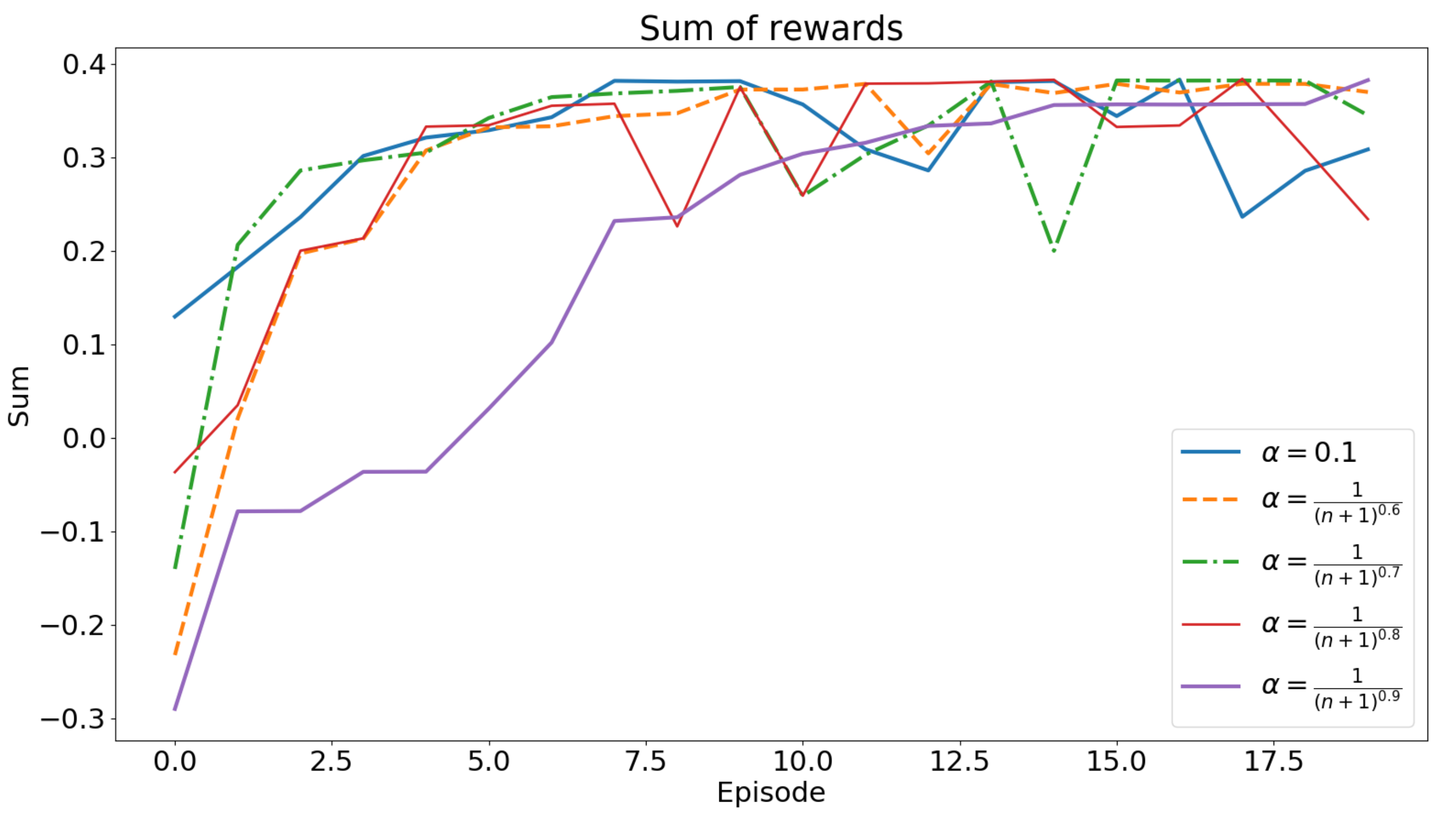}} 
       \caption{Roundabout training: Sum of rewards per episode.}
       \label{fig:test_round_sum}
\end{figure}

Using the algorithm in Figure \ref{fig:QL_alg}, the agent is trained to find the set of gains that minimize the error while the vehicle performs the two maneuvers. To choose the best set of gains, multiple training sessions were made, for each of the maneuvers, with different \textbf{$\pmb{\alpha}$} (learning rate) values. Figures \ref{fig:test_straight_sum} and \ref{fig:test_round_sum} show the sum of rewards of each episode (learning curve) of these trainings.

Each training session had 30 episodes for the lane changing maneuver and 20 episodes for the roundabout navigation. The training times of these tests were, on average, 24 hours  and 33 hours, respectively. Both figures show a convergence of the learning curves, which implies the success of the algorithm.


To define the set of gains for each maneuver, we used the set of gains that was picked more times throughout all the different $\alpha$ values: $\pmb{(3,21,21,0.7)}$ for the lane changing maneuver and $\pmb{(3.4,21,1,0.84)}$, for the roundabout maneuver. These are the sets of gains used in the validation tests.

For the validation process, the system performs each of the maneuvers with the corresponding chosen sets of gains. Then, the average Mean Square Error, $\overline{mse}$, of the trajectory position is calculated as 
\begin{equation}
\label{eq:mse}
\overline{mse} = \frac{1}{N} \sum_{i=1}^{N}{\frac{^be_{x_i}^2 + ^be_{y_i}^2}{2}},
\end{equation}
\noindent
where $N$ represents the number of data points registered. This process is repeated for different sets of gains spread through the range of gain values. The goal is to compare the performance of the chosen sets of gains with the performance of other sets of gains, while the system performs the maneuvers. Table \ref{tab:gain_stats} presents the average $\overline{mse}$ for the lane changing and the roundabout maneuver. For each set of gains, the system performs the maneuver 10 times, and then the highest average MSE registered is selected.

By default, CARLA does not consider any noise in the odometry sensor. To analyse the robustness of the system, noisy odometry measurements were simulated. Position noise is obtained by drawing random samples from a normal (Gaussian) distribution with a mean of 0 and a standard deviation of 0.1 meters. Orientation noise is obtained by drawing samples from the triangular distribution over the interval [-0.088, 0.088] rad and centered in 0. The third column in Table \ref{tab:gain_stats} shows $\overline{mse}$ and $\overline{mse_{\xi}}$, obtained under noisy conditions.

\begin{table}[h]
\caption{$\overline{mse}$ (without noise) and $\overline{mse_{\xi}}$ (with noise) for odometry measurements with different sets of gains, for the lane change and roundabout maneuvers.}
\centering
\begin{minipage}{0.47\linewidth}
\centering
\begin{tabular}{ccc}
\hline
\multicolumn{3}{c}{Lane Change} \\ \hline
\textbf{Gains} & $\pmb{\overline{mse}}$ & $\pmb{\overline{mse_{\xi}}}$ \\ \hline
$(0.1,1,6,0.7)$ & $6.94$ & $8.018$ \\
$(0.68,21,21,0.77)$ & $2.01$ & $6.02$ \\
$(1.26,6,11,0.84)$ & $1.628$ & $5.882$ \\
$(3,21,16,0.7)$ & $1.428$ & $5.591$ \\
$\pmb{(3,21,21,0.7)}$ & $\pmb{1.359}$ & $\pmb{5.589}$ \\
$(3,21,21,0.98)$ & $1.399$ & $5.637$ \\ \hline
\end{tabular}
\end{minipage}
\begin{minipage}{0.47\linewidth}
\centering
\begin{tabular}{ccc}
\hline
\multicolumn{3}{c}{Roundabout Navigation} \\ \hline
\textbf{Gains} & $\pmb{\overline{mse}}$ & $\pmb{\overline{mse_{\xi}}}$ \\ \hline
$(2.2, 21, 1, 0.98)$ & $0.245$ & $1.374$ \\
$(2.2, 16, 21, 0.77)$ & $0.230$ & $1.363$ \\
$(3.4,11,21,0.84)$ & $0.214$ & $1.388$ \\
$\pmb{(3.4,21,1,0.84)}$ & $\pmb{0.208}$ & $\pmb{1.347}$ \\
$(3.4, 21, 11,0.77)$ & $0.216$ & $1.385$ \\
$(4.6, 6, 1, 0.84)$ & $0.265$ & $1.429$ \\ \hline
\end{tabular}
\end{minipage}
\label{tab:gain_stats}
\end{table}

\begin{figure}[htbp]
\centering
\begin{minipage}{.5\textwidth}
    \centering
    \includegraphics[width=1\textwidth]{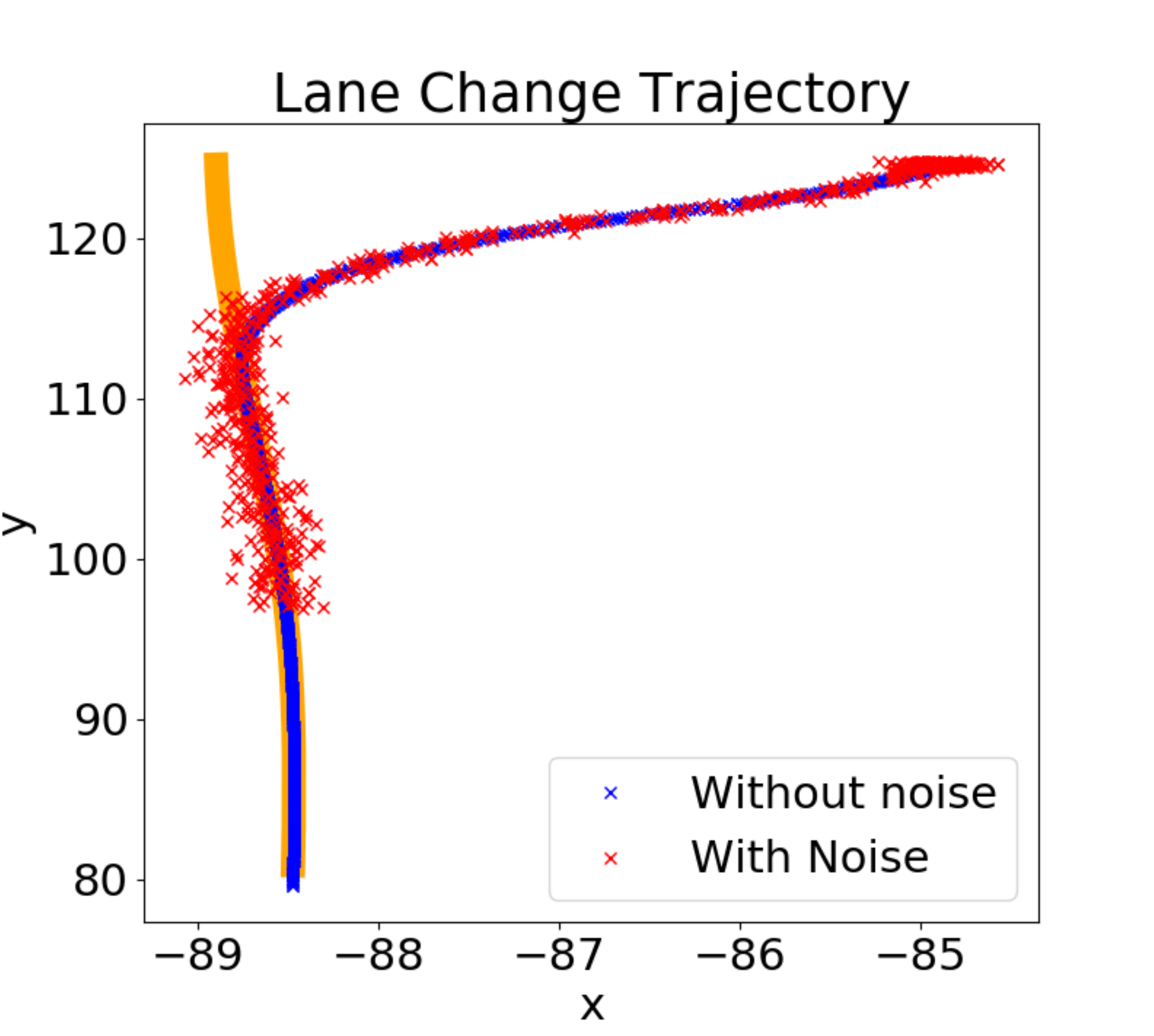}

    \includegraphics[width=1\textwidth]{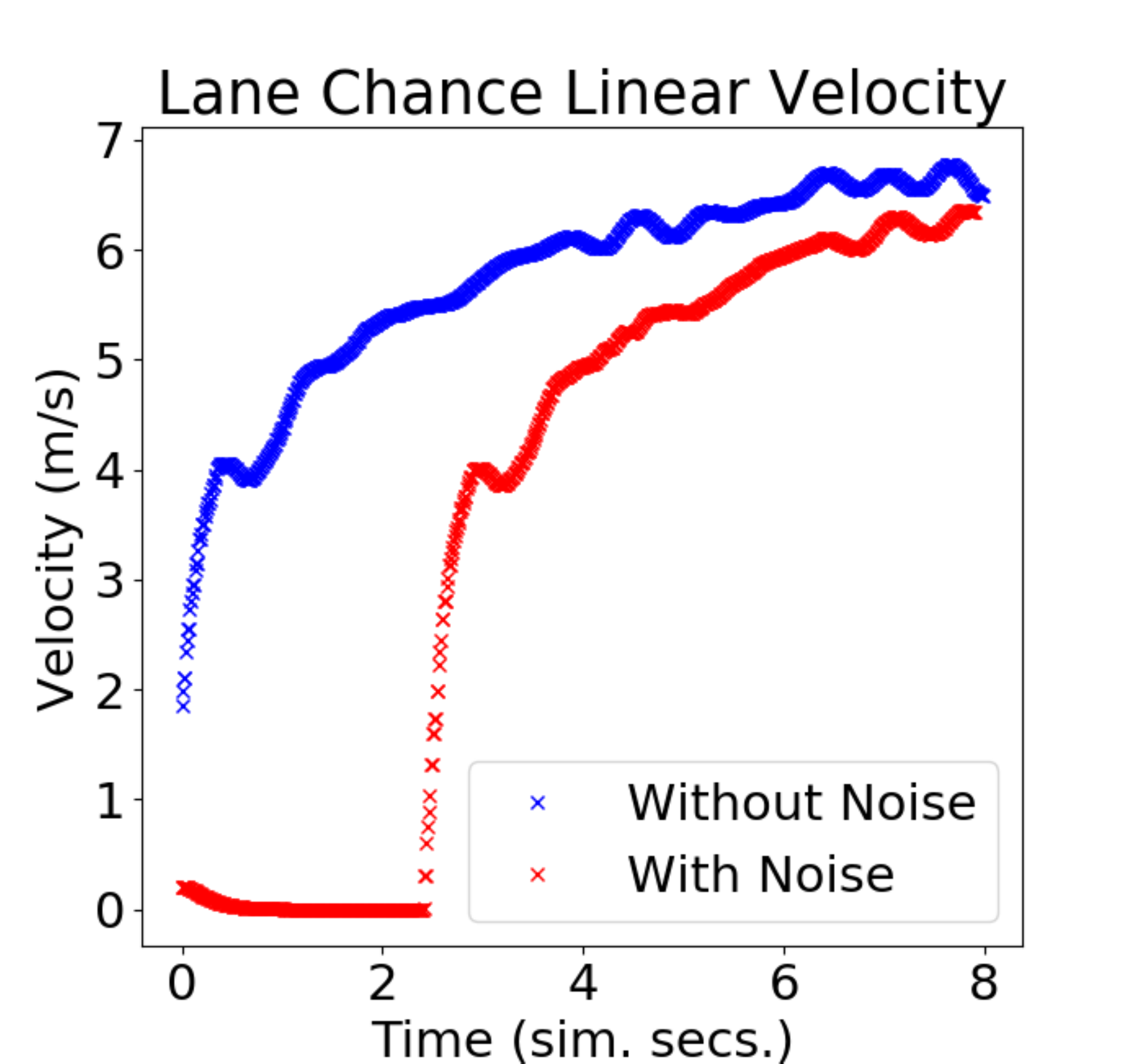}
    
    \includegraphics[width=1\textwidth]{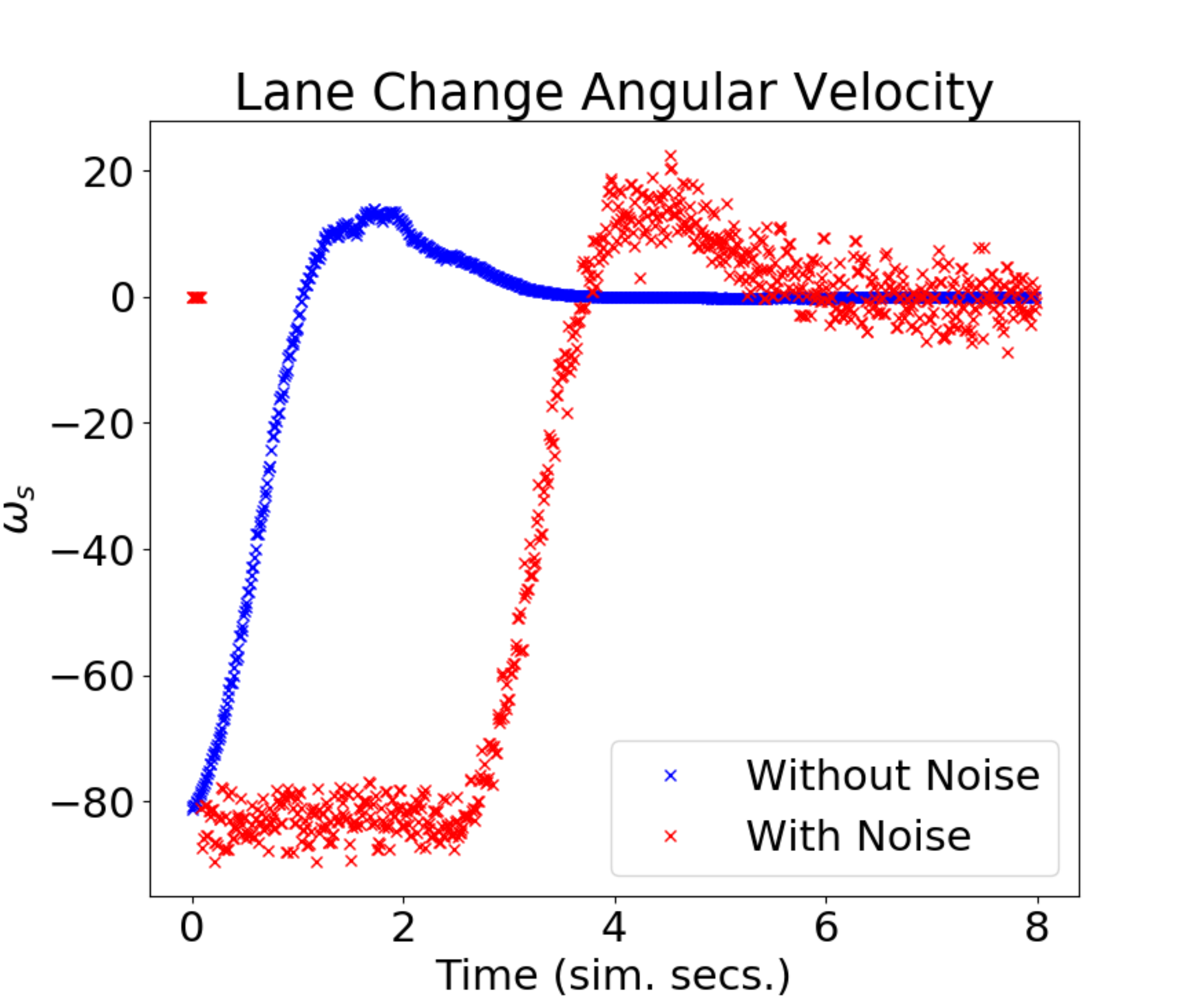}
    \caption{Lane changing Validation Test: trajectory, linear velocity, $v$ and angular velocity, $\omega_s$.}
    \label{fig:test_straight_sys_traj}
\end{minipage}%
\begin{minipage}{.5\textwidth}
    \centering
    \includegraphics[width=1\textwidth]{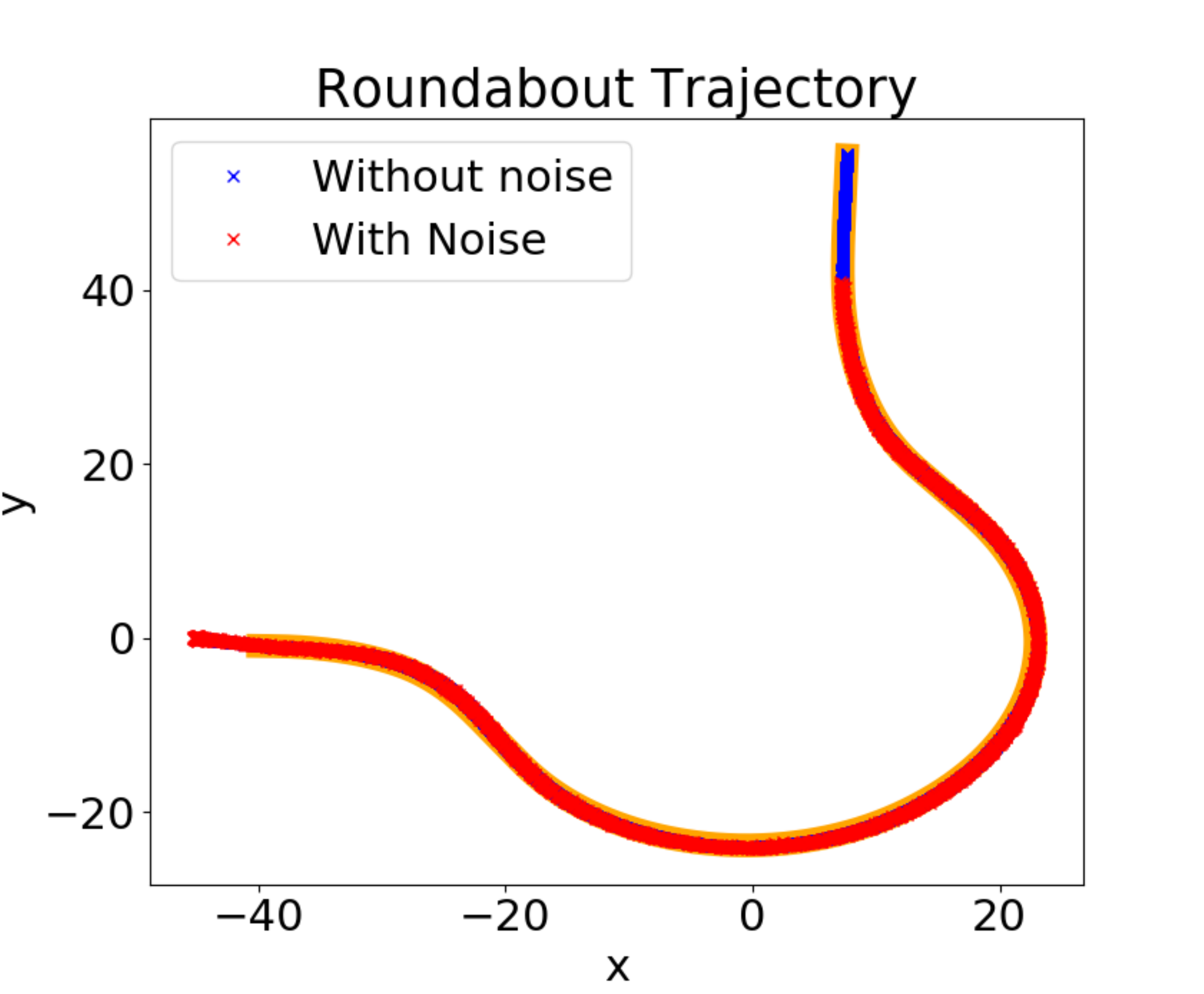}

    \includegraphics[width=1\textwidth]{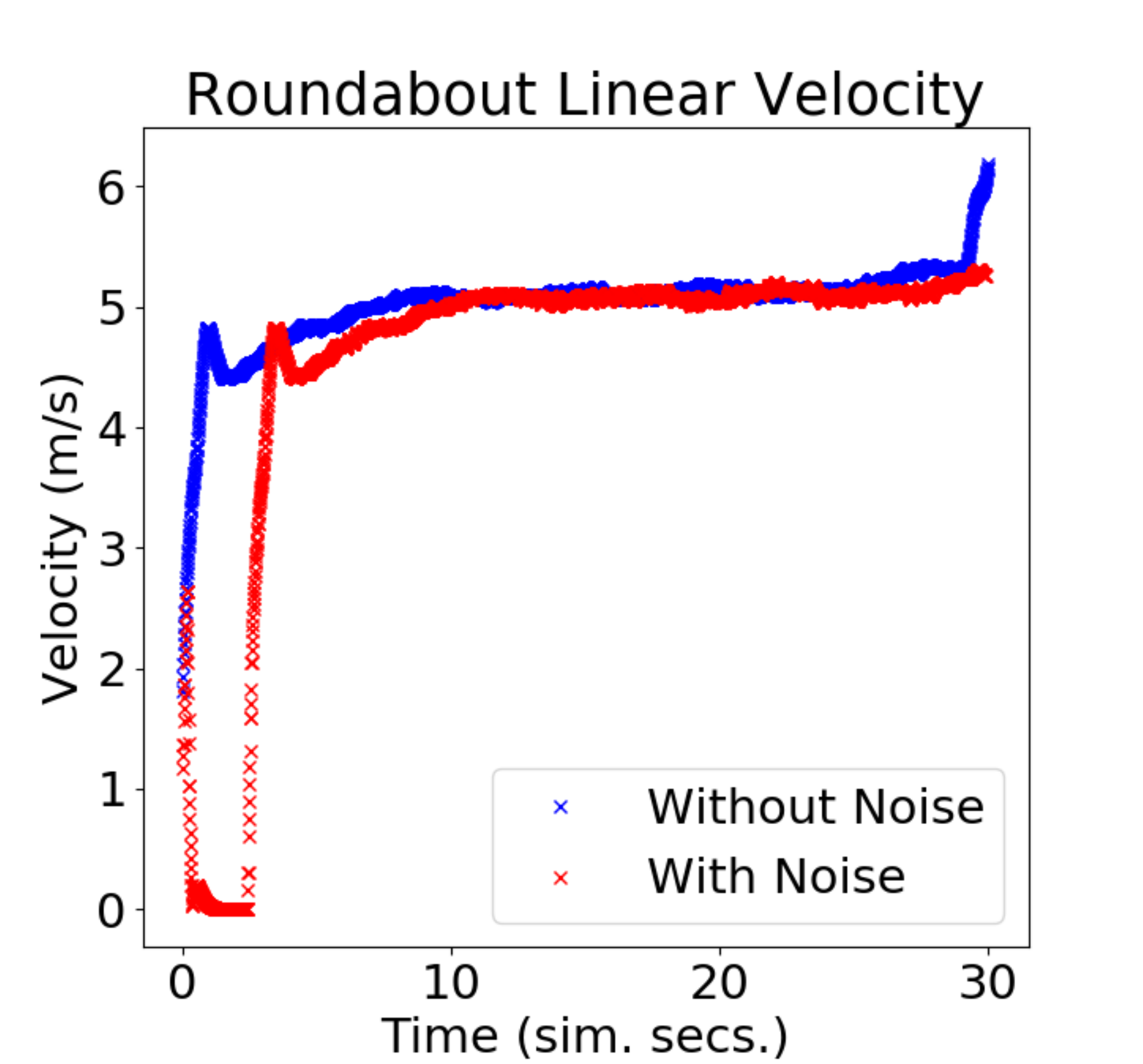}
    
    \includegraphics[width=1\textwidth]{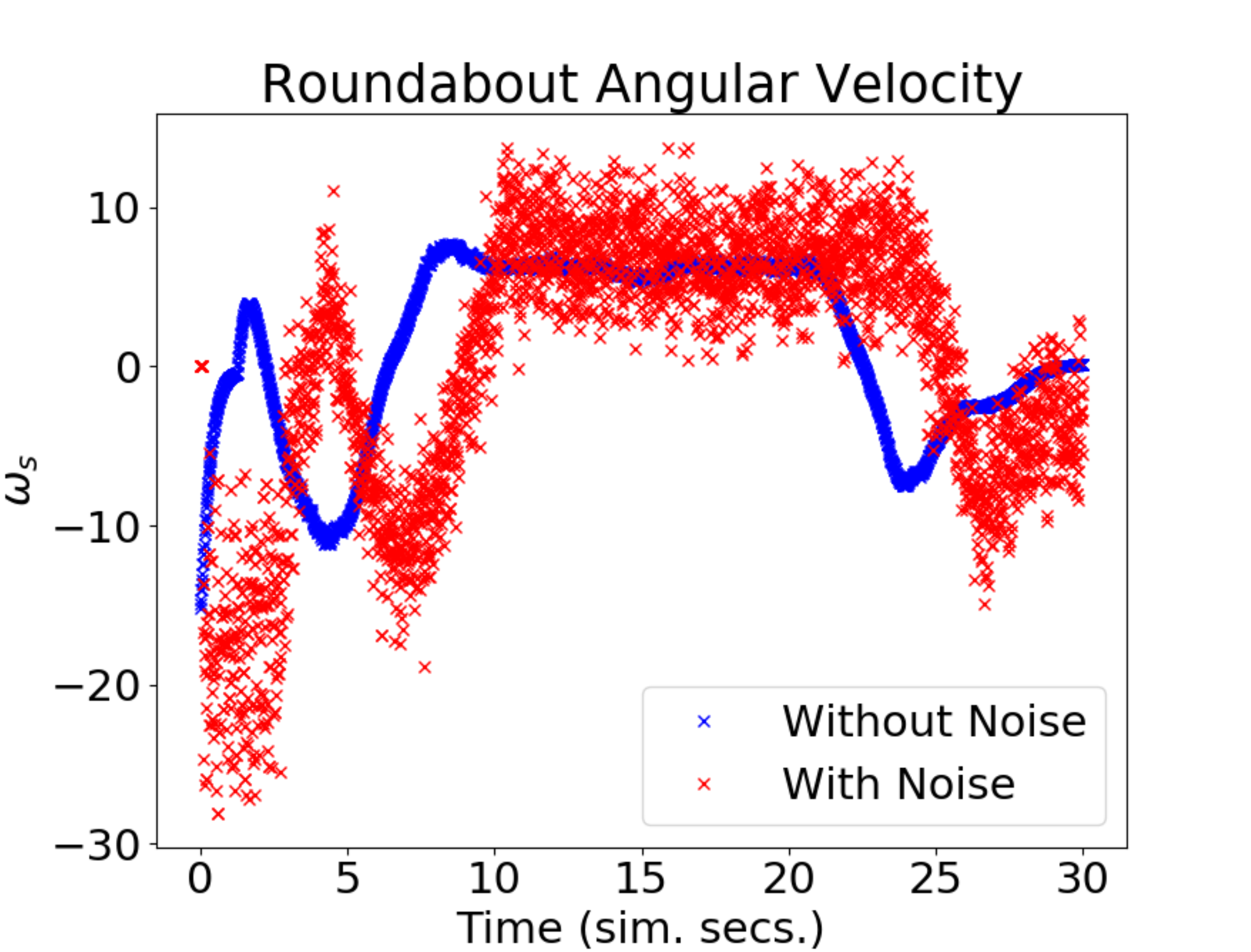}
    \caption{Roundabout Validation Test: trajectory - the reference and the trajectory are superimposed - linear velocity, $v$ and angular velocity, $\omega_s$}
    \label{fig:test_round_sys_traj}
\end{minipage}
\end{figure}

Figures \ref{fig:test_straight_sys_traj} and \ref{fig:test_round_sys_traj} show the trajectory, linear and angular velocity of the vehicle, without noise (blue) and with noise (red), and the reference path in orange, for the chosen gains. In the trajectory graph, the lengths of the trajectories with and without noise differ. This is because the duration of each validation test is fixed, and it takes longer for the system, with noisy odometry measurements, to take off. The delay in velocities, shown in the graphs below, corroborate this.
For both maneuvers,these figures and Table \ref{tab:gain_stats} show small lateral errors and $\overline{mse}$ values. A qualitative analysis of the values from Table \ref{tab:gain_stats} reveal that the gains chosen by the RL agent present the lowest $\overline{mse}$, implying that the chosen gains are in the neighbourhood of the values that minimize the trajectory error. Furthermore, comparing $\overline{mse}$ to $\overline{mse_{\xi}}$, it is possible to verify the system's robustness to some noise in the odometry sensor measurements, in the sense that the chosen gains continue to produce the lowest $\overline{mse}$ values.

\begin{figure*}[h]
\centering
\begin{minipage}{.55\textwidth}
    \centering
    \includegraphics[width=0.90\textwidth]{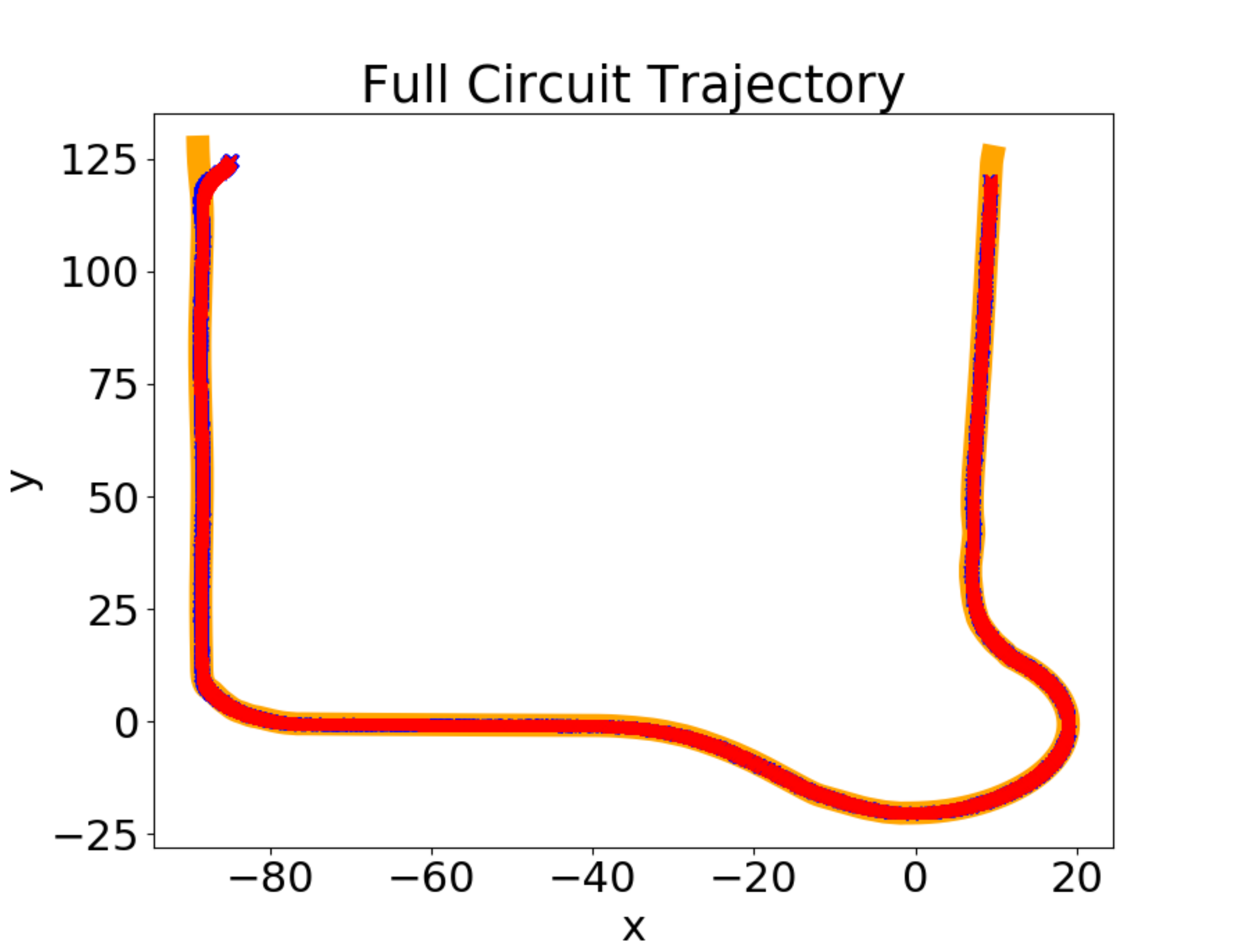}
    \caption{Full Circuit test: Trajectory \textendash~the reference and the trajectories are superimposed.}
    \label{fig:full}
\end{minipage}%
\begin{minipage}{.5\textwidth}
    \centering
    \includegraphics[width=0.9\textwidth]{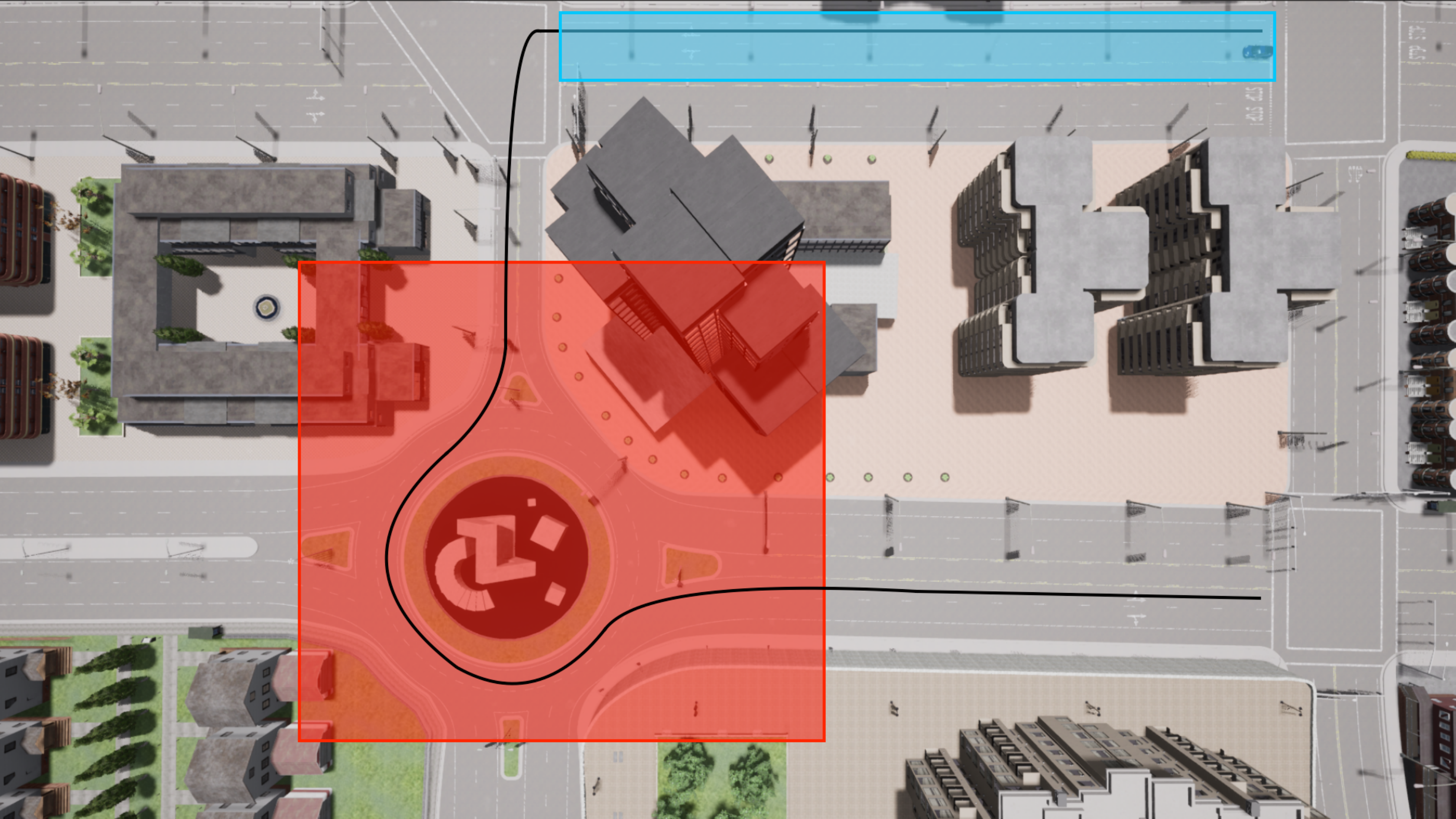}
    \caption{Organization of the simulation map in different zones.}
  \label{fig:town03_zones}
\end{minipage}
\end{figure*}

The system was also tested while navigating in the environment illustrated in Figure \ref{fig:town03_zones}, following the reference path defined in red, which included both maneuvers and a sharp left turn. The chosen gains for the blue and red zones were, respectively, $\pmb{(3,21,21,0.7)}$ and $\pmb{(3.4,21,1,0.84)}$. For testing purposes, the speed limit imposed is $4$ m/s. The results are presented in Figure \ref{fig:full}. It shows the trajectory performed by the system, without noise (blue) and with noise (red), and the reference path, in orange, which are, on average, superimposed: the system successfully follows the reference path, without any major errors or collisions.
\begin{table}[h]
\caption{$\overline{mse}$ and $\overline{mse_{\xi}}$ for different sets of gains, for the full circuit.}
\centering
\begin{tabular}{ccc}
\hline
\textbf{Gains} & $\pmb{\overline{mse}}$ & $\pmb{\overline{mse_{\xi}}}$ \\ \hline
$(1.84, 1, 1, 0.91), (2.2,6,11,0.7)$ & $1.213$ & $1.32$ \\
$\pmb{(3, 21, 21, 0.7)}, \pmb{(3.4, 21, 1, 0.84)}$ & $\pmb{0.181}$ & $\pmb{0.363}$ \\
$(3, 21, 21, 0.98), (5.8, 16, 11, 0.84)$ & $0.185$ & $0.673$ \\ \hline
\end{tabular}
\label{tab:full_stats}
\end{table}
Table \ref{tab:full_stats} presents $\overline{mse}$ and $\overline{mse_{\xi}}$ values for different sets of gains, with the reference path defined in Figure \ref{fig:town03_zones}. As with the maneuvers, the results show that the chosen sets of gains produce the lowest $\overline{mse}$ out of a wide range of gains, which suggests the proximity of the chosen gains to the optimal gain values. 

\section{An argument on dependability}

This section aims at sketching a framework to research dependability properties in the RL enabled autonomous car
setting.

The GNC architecture, typical of a wide class of robotics systems, fully applies
to the context of autonomous cars.
The Control block accommodates multiple controllers tuned to specific driving conditions
and the Guidance block selects which of the controllers is used at every instant.
Switching between controllers may be required in several situations, e.g.,
overtaking maneuvers, or changing between straight and twisty roads or even changing between smooth and aggressive driving styles, and hence the overall system is of hybrid nature.
Often, the switching mechanism will have the form of a finite state machine and the overall Control block can
be described as an affine model,

\begin{equation}
\label{eq:controller_structure}
C = D + C_1 u_1 + C_2 u_2 + \ldots + C_n u_n
\end{equation}

\noindent
where $D, C_1, C_2, \ldots, C_n$ can be assumed smooth vector fields representing each a controller,
and $u_1, u_2, \dots, u_n$ stand for the switching control variables which are 0 whenever their respective controller
is not active and $D$ is an affine term which may represent a controller terms that must be always present (and hence
not subject to any sudden change of structure).


In general, a regular mission, i.e., driving between any two locations, includes moving along the same lane,
changing lanes, moving through crossings and roundabouts, among other more specific maneuvers such as parking.
As it is well known from Control theory, switching among stable systems may lead to instability (see for instance
 \cite{Lin.Antsaklis:2009} in the framework of switched linear systems).

Let $Q_1, Q_2, \ldots, Q_n$ be the accumulated reward trajectories for a set of $n$ controllers trained using RL.
For each individual maneuver, the complete trajectory of the reward obtained can be assumed known from the training process.
The value of an accumulated reward at the final of the training is an indicator of the quality of the policies
found (\cite{BartoSutton}, pp 54-55), if the policies are allowed to run for a time long enough (so that they can
reach their goals) and one can assume them globally exponentially stable.

Using the converse Lyapunov theorem, this also means that there are Lyapunov functions $V_1, V_2, \ldots, V_n$, associated
with each of the individual
sets of controller parameters, which, surely, have derivatives $DV_1 < 0, DV_2 < 0, \ldots, DV_n <0$.
Therefore, one can compose a candidate to Lyapunov function as

\begin{equation}
\label{eq:lyapunov_candidate}
V = V_1 u_1 + V_2 u_2 + \ldots + V_n u_n
\end{equation}

\noindent
where $u_1, u_2, \ldots, u_n$ as defined above.

This technique has been reported in the literature when the $V_i$ are quadratic functions and the $C_i$ are polynomial
vector fields (see for instance \cite{Tan.Packard:2004}, \cite{Papachristodoulou.Prajna:2002}).
In this paper we aim at a more general approach.
The system formed by the finite state machine structure used to switch among controllers, composed with the controllers
and the car (assumed to be a regular kinematic structure such as the well known car-like robot)
can be shown to be upper semicontinuous (USC). 
Writing \eqref{eq:controller_structure} in the alternative set-valued map form as,
$
C = \cup_{i=1}^n C_i u_n,
$
following the definition of USC set-valued map (see for instance, Definition 1 in \cite{Aubin.Cellina:1984}, p. 41,
or \cite{Smirnov:2001}, pp. 32-33), the overall system is USC as they have closed values and, by
Proposition 2 in \cite{Aubin.Cellina:1984} this means that the corresponding graphs are closed.
and hence one is in the conditions required by
the generalized Lyapunov theorem in \cite{Aubin.Cellina:1984}, for asymptotic stability, 

\begin{equation}
\label{eq:lyapunov_generalized}
D_{+}V(x) < -W(x)
\end{equation}

\noindent
with $W$ a strictly positive monotonic decreasing function and $D_{+}V(x)$ representing the contingent derivative
of $V$ at $x$, and $V$ lower semi-continuous, and an equilibrium can be reached.

In general, in the car control context, the switching will occur at arbitrary instants, though a minimal separation
between switching instants can safely be assumed
(as in a realistic situation a car will not switch arbitrarily fast between behaviours in a repetitive way).
Also, the switching will make $V$ to have bounded discontinuities (at switching instants)
and before each discontinuity will have a monotonic decreasing trend (as each behavior is assumed asymptotically stable) 
 and hence \eqref{eq:lyapunov_candidate} can be safely assumed to be lower semi-continuous.

The $Q_i$ values are known {\em a priori} from the training phase and can be monitored while in real conditions and be
constantly compared with the training, this yielding a performance metric that can be used for control purposes,
namely defining thresholds to control the switching (switch only if currently observed $Q_i$ is close enough to the
value recorded during training).
This ensures that there exists an envelop function $W$ such that  \eqref{eq:lyapunov_generalized} holds.

\section{Conclusions}
\label{sec:conclusion}

This paper proposes a RL-based  path tracking control system for a four-parameter architecture. The tuning is done by an \textit{educated} Q-Learning algorithm, that minimizes the lateral and steering trajectory errors of the vehicle while performing lane change and
 roundabout maneuvering.

The \textit{educated} Q-Learning variant introduced in the paper uses a reduced action space during training, allowing for a faster convergence to the final set of gains (though it can lead to sub-optimal solutions).

The trajectories in Figures \ref{fig:test_straight_sys_traj}, \ref{fig:test_round_sys_traj}, and \ref{fig:full}, as well as the velocity values registered during these experiments demonstrate that the system does not engage in unsafe behaviour, like collisions or excessive velocity. It also consistently follows the reference with small error. The MSE values in Table \ref{tab:gain_stats} suggest that the algorithm can efficiently tune the gains to values that are in the neighbourhood of the gain values that minimize the trajectory errors.

Literature on the subject of autonomous driving control is populated with Neural Network  (NN) based controllers in the role of supervisors
of the tracking function (see for instance \cite{Fenyes.et.al:2021} on NNs to match the car model and subsequent robust control design).
For the proposed architecture, the argument on dependability developed in the previous section shows that the overall system
has a stability property (which amounts to safe driving). 

Future work includes the (i) refinement of the dependability argument to account for the stochastic nature of the $Q_i$ values, 
(ii) tests under noisy/uncertain conditions, and (iii) trials in a real vehicle.

\section*{Acknowledgements}
This work was partially supported by project LARSyS-FCT Project UIDB/\-
50009/2020.

%

\end{document}